\documentclass[12pt,preprint]{aastex62}

\begin{document}

\title{Towards an Understanding of the Massive Red Spiral Galaxy Formation}

\correspondingauthor{Cai-Na Hao}
\email{hcn@bao.ac.cn}

\author{Rui Guo}
\affiliation{Tianjin Astrophysics Center, Tianjin Normal University, Tianjin 300387, China}

\author[0000-0002-0901-9328]{Cai-Na Hao}
\affiliation{Tianjin Astrophysics Center, Tianjin Normal University, Tianjin 300387, China}

\author{Xiaoyang Xia}
\affiliation{Tianjin Astrophysics Center, Tianjin Normal University, Tianjin 300387, China}

\author[0000-0002-8614-6275]{Yong Shi}
\affiliation{School of Astronomy and Space Science, Nanjing University, Nanjing 210093, China}
\affiliation{Key Laboratory of Modern Astronomy and Astrophysics, Nanjing University, Ministry of Education, Nanjing 210093, China}

\author{Yanmei Chen}
\affiliation{School of Astronomy and Space Science, Nanjing University, Nanjing 210093, China}
\affiliation{Key Laboratory of Modern Astronomy and Astrophysics, Nanjing University, Ministry of Education, Nanjing 210093, China}

\author{Songlin Li}
\affiliation{School of Astronomy and Space Science, Nanjing University, Nanjing 210093, China}
\affiliation{Key Laboratory of Modern Astronomy and Astrophysics, Nanjing University, Ministry of Education, Nanjing 210093, China}

\author{Qiusheng Gu}
\affiliation{School of Astronomy and Space Science, Nanjing University, Nanjing 210093, China}
\affiliation{Key Laboratory of Modern Astronomy and Astrophysics, Nanjing University, Ministry of Education, Nanjing 210093, China}

\begin{abstract}

To understand the formation and quenching processes of local massive red spiral
galaxies with $M_{\ast} > 10^{10.5}M_{\odot}$, we perform a statistical
analysis of their spectroscopic and structural properties, and compare them
with elliptical and blue spiral galaxies of similar mass. The sample was
selected from the stellar mass catalog of galaxies in SDSS DR7, according to
their locations on the $u-r$ color-stellar mass diagram. We find that red
spirals harbor compact cores with high stellar mass surface densities measured
by $\Sigma_1$ and they are bulge-dominated. Particularly, the red spirals,
especially their bulges follow the $\Sigma_1$-$M_{\ast}$ ridgeline for quenched
galaxies. Furthermore, the red spirals show similarly large central
D$_n(4000)$, high [Mg/Fe] and dark matter halo mass to ellipticals. These
results suggest that the bulges of red spirals formed within a short timescale
before redshift $\sim1-2$ and were quenched via a fast mode, similar to
ellipticals. Careful examinations of the optical morphologies reveal that $\sim
70\%$ of red spirals show strong bars, rings/shells and even merging features,
which suggests that interactions or mergers might have played an important role
in the formation of red spirals. In contrast, most of the massive blue spirals
have completely different spectral and structural properties from red spirals.
However, the blue spirals with high $\Sigma_1$ ($\Sigma_1 > 10^{9.5} M_\odot \, {\rm
kpc}^{-2}$) show similar structural and morphological properties, as well as
similar halo mass and HI mass to red spirals. We discuss
rejuvenation from red to blue as a possible explanation for these high
$\Sigma_1$ blue spirals.

\end{abstract}


\section{INTRODUCTION}

Since the discovery of the galaxy color bimodality in the color-magnitude or
color-stellar mass diagrams both locally and at high redshifts
\citep[e.g.,][]{Kauffmann2003b,Baldry2004,Baldry2006,Bell2004,Faber2007,Ilbert2010},
there have been mounting works investigating the evolutionary pathways from the
blue cloud of star-forming galaxies to the red sequence of quiescent galaxies
\citep[e.g.,][]{Bell2004,Faber2007,Marchesini2014,Schawinski2014}.  However,
when the morphologies of galaxies are taken into account, the bimodality almost
disappears \citep{Schawinski2014}. This is a result of the close relation
between galaxy colors and morphological types: early-type galaxies are
mainly located in the red sequence, while disk galaxies mostly populate the
blue cloud. Therefore, the popular picture proposed for galaxy evolution is
that quenching processes are accompanied with structure transformation, i.e.,
quenched massive spheroidals were transformed from blue star-forming disk
galaxies.  On the other hand, there is a striking feature in the color-stellar
mass diagram for spiral galaxies: while the less massive spirals ($M_{\ast}<
3\times10^{10} M_{\odot}$) occupy the blue cloud region, a population of
massive spiral galaxies ($M_{\ast}> 3\times10^{10} M_{\odot}$) are in the red
sequence \citep{Schawinski2014}. The existence of such massive red spiral
galaxies challenges the scenario that galaxy quenching must be in company with
the morphological transformation \citep[e.g.,][]{Skibba2009, Bundy2010,
Masters2010, Fraser2018}.

It has been over 40 years since the first studies for passive spiral galaxies
\citep[e.g.,][]{vandenBergh1976, Dressler1999, Poggianti1999,
Goto2003,Skibba2009}.  At earlier times, the interests were mainly on the
environmental effects on the formation of red spirals. For galaxies in
clusters, ram-pressure could strip off the gas in and around galaxies and hence
shut down the star formation.  In the past decade, with the advent of several
wide or deep photometric and spectroscopic surveys, such as Sloan Digital Sky
Survey \citep[SDSS,][]{York2000}, Cosmic Evolution Survey
\citep[COSMOS,][]{Scoville2007}, {\em Galaxy Evolution Explorer} \citep[{\em
GALEX},][]{Martin2005} and {\em Wide-field Infrared Survey Explorer}
\citep[{\em WISE},][]{Wright2010}, passive spirals have attracted more
attention. Many efforts have been invested in understanding the origin of red
spirals by studying their stellar populations, structures and environments
\citep[e.g.,][]{Bundy2010, Masters2010, Robaina2012, Tojeiro2013, Fraser2018}.
Based on the Galaxy Zoo project (GZ), \citet{Masters2010} found that at high
stellar masses ($M_* > 10^{10} M_{\odot}$), a significant fraction of spirals
are red and the environment is not sufficient to quench these massive spirals.
In consideration of the old stellar populations hosted by red spirals
\citep[see also][]{Robaina2012, Tojeiro2013} and the intact disk morphology,
\citet{Masters2010} proposed that red spirals might be old spirals that have
exhausted all of their gas. Meanwhile, \citet{Bundy2010} investigated the
evolution of passive spirals since $z$$\sim$1-2 based on the COSMOS survey and
made extensive discussions about possible origins of red spirals. They found
that red spirals have more concentrated light distribution than blue spirals,
and hence red spirals are unlikely to originate from blue spirals. It is still
unclear how red spirals formed and the star formation was quenched.

Galaxy bulges are prominent components of massive spiral galaxies.
\citet{Robaina2012} compared the central stellar population properties of
massive ($M_* > 10^{10.4} M_{\odot}$) red spirals with elliptical galaxies based
on SDSS Data Release 7 (DR7) and GZ. They found that the formation epoch and the star formation
duration are related to the bulge mass. Interestingly, bulge building is also a
key process in galaxy quenching \citep{Martig2009, Bluck2014}. Based on half
million local SDSS galaxy sample, \citet{Bluck2014} systematically investigated
the connection of quenching mechanisms to galaxy properties and found that the
bulge mass is the dominator of the passive galaxy fraction, indicating the
importance of morphological quenching. Therefore, the
bulge is an important component in our understanding of galaxy formation and
evolution. Several parameters have been used to characterize bulge properties,
such as bulge mass, bulge-to-total light/mass ratio, bulge surface mass
density, sersic index etc. More recently, the stellar mass surface density
within a radius of 1\,kpc ($\Sigma_1$), a measure of the innermost structure of
galaxies, has been considered as a more powerful probe of galaxy quenching
\citep{Cheung2012,Fang2013,Barro2017a}. The scale of 1\,kpc is coincident with
those of the young bulges forming at $z\sim2$, as revealed by the ALMA and HST
observations \citep{Barro2016,Barro2017b,Tadaki2017a,Tadaki2017b,Newman2018}.  
Taken together, $\Sigma_1$ is linked with both bulge formation and galaxy quenching.
A visit of this parameter will help to understand these two processes.

Many quenching mechanisms have been explored in the literature. Apart from the
aforementioned morphological quenching, halo quenching
\citep{DekelBirnboim2006}  is also among the most popular ones for central
galaxies.  It is expected that galaxies hosted by dark matter halos above some
critical halo mass are unable to form new stars due to the shock heating of
circumgalactic gas. This critical halo mass is about $10^{12} M_{\odot}$
\citep{DekelBirnboim2006}.  \citet{Fraser2018} examined a sample of 35 nearby
passive spiral galaxies that consists of 30 massive ones with $M_{\ast}>
1\times10^{10} M_{\odot}$.  After investigating the bar fractions and
environments, they concluded that the quenching mechanisms for massive passive
spiral galaxies are still a puzzle.

In this work, we concentrate on a relatively large sample of massive ($M_{*} >
10^{10.5} M_{\odot}$) red spiral galaxies and compare them to blue spiral and
red elliptical galaxies above the same mass limit. By investigating their
central stellar population properties, detailed morphological features,
$\Sigma_1$, bulge properties, gas contents and dark matter halo masses etc, we
expect to shed light on the formation and quenching mechanisms of red massive
spiral galaxies. This is the first paper in this paper series. In a followup
study by \citet{Hao2019}, based on Mapping Nearby Galaxies at the Apache Point
Observatory (MaNGA) \citep{Bundy2015} two-dimensional spectra, we explored the spatially resolved
stellar population and kinematical properties using subsamples of galaxies in
this work. In Section 2, we describe the sample selection and parameter
derivation for our sample galaxies.  We present the results in Section 3. In
Section 4 and 5, we discuss and summarize our findings, respectively.
Throughout this paper, we adopt the \citet{Chabrier2003} initial mass function
(IMF) and a cosmology with $H_{\rm 0}=70\,{\rm km \, s^{-1} Mpc^{-1}}$,
$\Omega_{\rm m}=0.3$ and $\Omega_{\rm \Lambda}=0.7$.

\section{SAMPLE AND PARAMETERS}

\subsection{Sample Selection\label{subsec:sample}}

Our samples were drawn from the catalog of \citet{Mendel2014} by constraining
the redshift range to $0.02 < z < 0.05$, and further requiring the luminosity
of the galaxies in the range of $M_{z,\rm Petro} < -19.5$ mag
\citep{Schawinski2014} in order to derive the distribution of a parent sample
covering wider range of stellar mass in the color-stellar mass diagram. For our
working samples, we only selected galaxies with the total stellar mass
$M_{\ast} > 10^{10.5} M_{\odot}$. This yielded a sample of 11,172 massive
galaxies. We will use spirals and ellipticals to denote our massive spiral and
elliptical galaxies, hereafter.

We then used the visual morphological classifications from the GZ 1
project \citep{Lintott2008, Lintott2011} to obtain the morphological types of
the galaxies. This resulted in 3,908 spirals and 3,261 elliptical galaxies. To
ensure the reliability of the photometric bulge-disk decomposition and minimize
the dust-reddening effect on the color measurements for spiral galaxies,
the spiral galaxies with minor-to-major axis ratio $b/a < 0.5$ were excluded.  In
addition, we examined each image of these spiral galaxies to further remove
false face-on galaxies\footnote{As pointed out by \citet{Gadotti2009}, axial
ratio derived from SDSS are not correct in some cases.}. This step reduced the
number of face-on spiral galaxies to 1,914.

\begin{figure}
\plotone{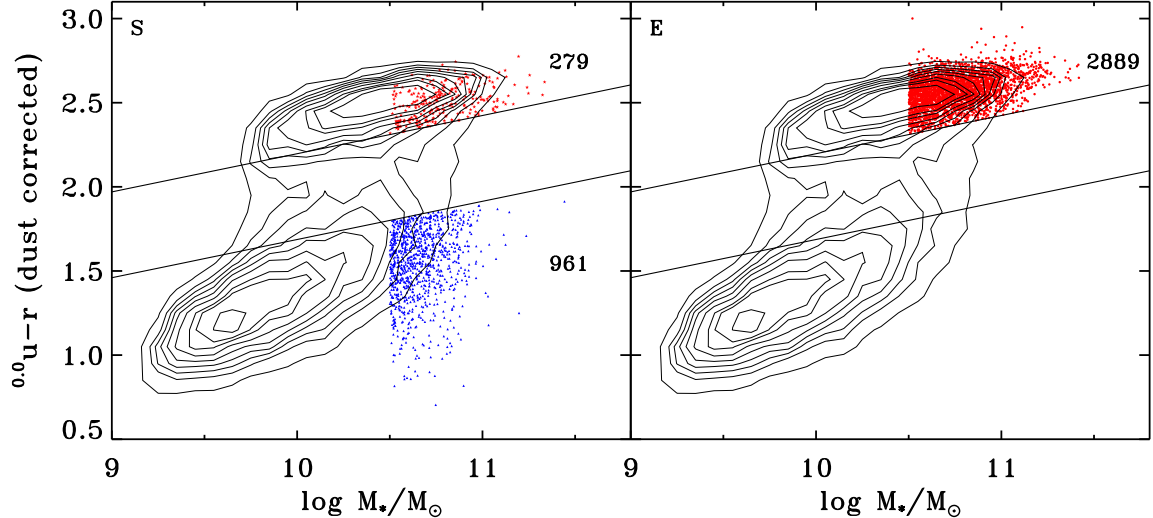}
\caption{Dust-corrected {\em u-r} color-stellar mass diagram for massive
red and blue spiral ({\em left}) and red elliptical ({\em right}) galaxies.
Blue triangles and red stars represent spirals galaxies
located in the red sequence and blue cloud, respectively.
Red circles represent elliptical galaxies located in the red sequence.
The number of each type of our sample galaxies is shown on the
right side of each panel.
The contours show the number density distribution of a parent sample of galaxies
in the redshift range of $0.02 < z < 0.05$ and absolute $z-$band magnitude
of $M_{z,\rm Petro} < -19.5$ in the catalog of \citet{Mendel2014}.
The black solid lines denote the boundaries of the green valley defined by
the number density distribution of the parent sample galaxies.}
\label{sample.ps}
\end{figure}

Finally, we utilized the dust-corrected {\em u-r} color-stellar mass diagram to
single out blue cloud galaxies and red sequence galaxies (see Figure
\ref{sample.ps}), using similar criteria of \citet{Guo2016}. This selection
produced 279 red, 961 blue spiral galaxies and 2,889 red elliptical galaxies as
our working samples. This shows that more than ten percent of massive spiral
galaxies (279/1,914) are red, and $50\%$ of them are blue.  We note that our
samples are not complete, but should be representative samples for galaxies in
each category. Especially, when some parameters are not available for a subset
of our sample galaxies (see Section \ref{subsec:data}), only the subsamples
with available measurements will be used.

\begin{figure}
\plotone{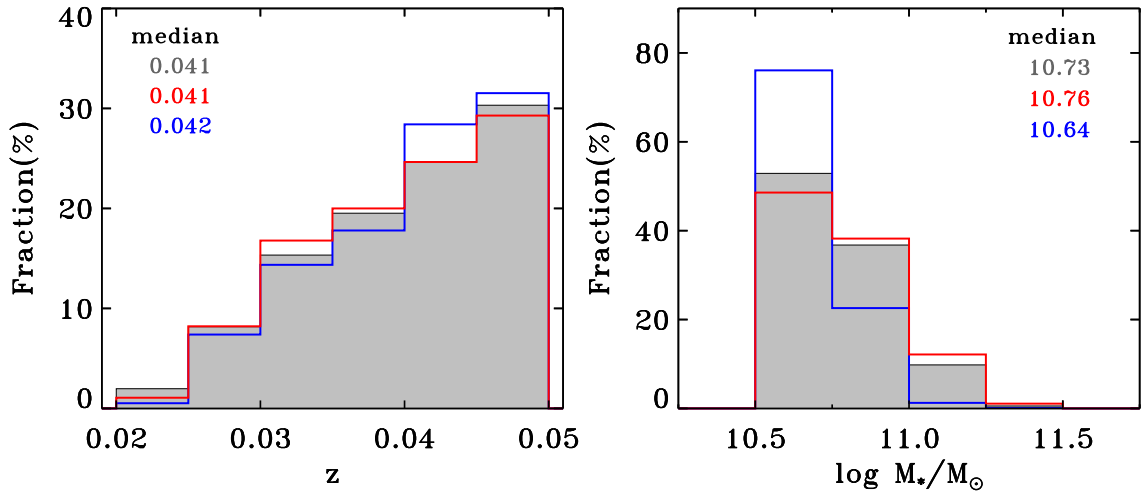}
\caption{Redshift ({\em left}) and stellar mass ({\em right}) distributions
for our sample galaxies.
The red and blue empty histograms represent spirals belonging to the
red sequence and blue cloud, respectively.
The gray filled histogram is for ellipticals in the
red sequence.
The median values for the respective
categories are labeled in each panel.}
\label{newlgmdistri.ps}
\end{figure}

Figure \ref{newlgmdistri.ps} shows the redshift and stellar mass distributions
for the three subsamples. It is clear from the left panel of Figure
\ref{newlgmdistri.ps} that the three samples have similar redshift
distributions. However, from the right panel of Figure \ref{newlgmdistri.ps},
although the distributions of stellar mass for the red spirals and ellipticals
are similar, quite a large fraction of blue spirals shows relatively lower
stellar mass, comparing with red spiral and elliptical galaxies.

Given that the color measurements of our sample galaxies may be influenced by
the presence of AGNs, we evaluate this effect by examining the optical spectral
types. The sample of \citet{Mendel2014} does not include Seyfert 1 galaxies.
We identified Seyfert 2 galaxies by widely used BPT diagrams proposed by
\citet{Baldwin1981} and developed by \cite{Kauffmann2003a} and
\citet{Kewley2001, Kewley2006}.  It turned out that only 4.3\% (12/279) red spirals, 4.1\%
(39/961) blue spirals and 0.4\% (11/2,889) ellipticals are Seyfert 2 galaxies.
Since the fractions of Seyfert 2 galaxies are small and would not affect our
statistical results, we did not remove them from our samples.

\subsection{Parameter Derivation\label{subsec:data}}

The parameters used in this work were mainly obtained from the public database,
except for $\Sigma_1$. We will briefly describe the derivation of the
parameters below, and refer the reader to the original papers for more details.

The bulge masses (M$_{\rm bulge}$), disk masses (M$_{\rm disk}$) and total
stellar masses were retrieved from \citet{Mendel2014}. The {\em u-} and {\em
r-}band magnitudes of the bulge components were from \citet{Mendel2014} (J. T.
Mendel, private communication) and \citet{Simard2011}, respectively.
\citet{Simard2011} performed photometric measurements on the {\em g-} and {\em
r-}band images of galaxies from the SDSS DR7 \citep{York2000,Abazajian2009}
using three sets of models: a single S{\'e}rsic profile, a de Vaucouleurs bulge
plus exponential disk, and a S{\'e}rsic bulge plus exponential disk.  A
probability parameter $P_{pS}$ was derived to judge the necessity of a
bulge+disk model compared to a pure S{\'e}rsic model, and $P_{pS} \le 0.32$ was
proposed as a criterion of real bulge+disk systems.  \citet{Mendel2014}
extended their work to the {\em u, i} and {\em z} bands and focused on
either a single S{\'e}rsic profile or a de Vaucouleurs bulge plus exponential
disk fitting to derive the bulge, disk and total stellar masses via SED
fitting.  They classified the galaxies into different types according to their
best-fit two-dimensional profiles.  We used the combination of the $P_{pS}$
value provided by \citet{Simard2011} and the best-fit profile types in
\citet{Mendel2014} to distinguish a genuine bulge+disk system from a single
profile system, and adopted the corresponding stellar mass derived from the
best-fitting profile. The stellar masses with dust corrections were used. For
galaxies with extremely red colors, the bulge masses are sometimes severely
overestimated with dusty models, so we used the dust-free results for galaxies
with M$_{\rm bulge}$ + M$_{\rm disk}$ higher than their total masses by 1
$\sigma$ or above, as recommended by \citet{Mendel2014}. There are 92.5\%
(258/279) red spirals and 87.4\% (840/961) blue spirals being best fitted with
a bulge+disk model, and only these galaxies will be used in the analysis of the
effective radius (R$_e$) and $u-r$ color for the bulge components, the bulge mass and the bulge-to-total
stellar mass ratio (B/T). The {\em u-} and {r-}band magnitudes of the bulge
components were corrected for internal extinctions using the E(B-V) provided by
the Oh-Sarzi-Schawinski-Yi (OSSY) catalog
\citep{Oh2011}\footnote{http://gem.yonsei.ac.kr/$\sim$ksoh/wordpress}.
The B/T was defined as M$_{\rm bulge}$/(M$_{\rm bulge}$+M$_{\rm disk}$). 

The dust-corrected {\em u-r} colors for the entire galaxies were calculated
from the {\em u-} and {\em r-}band model magnitudes that were retrieved from
the SDSS DR7. We first applied {\em k-}corrections to the {\em u-} and {\em
r-}band model magnitudes based on the New York University Value-Added Galaxy
Catalog \citep[NYU-VAGC;][]{BlantonRoweis2007}. Then we corrected the
{\em k}-corrected  {\em u-} and {\em r-}band model magnitudes for the foreground
Galactic extinctions using the dust maps from \citet{Schlegel1998}, and for the
internal dust extinctions using the E(B-V) from stellar continuum fitting
provided by the OSSY catalog, based on the \citet{Calzetti2000} extinction law.

\begin{figure}
\plotone{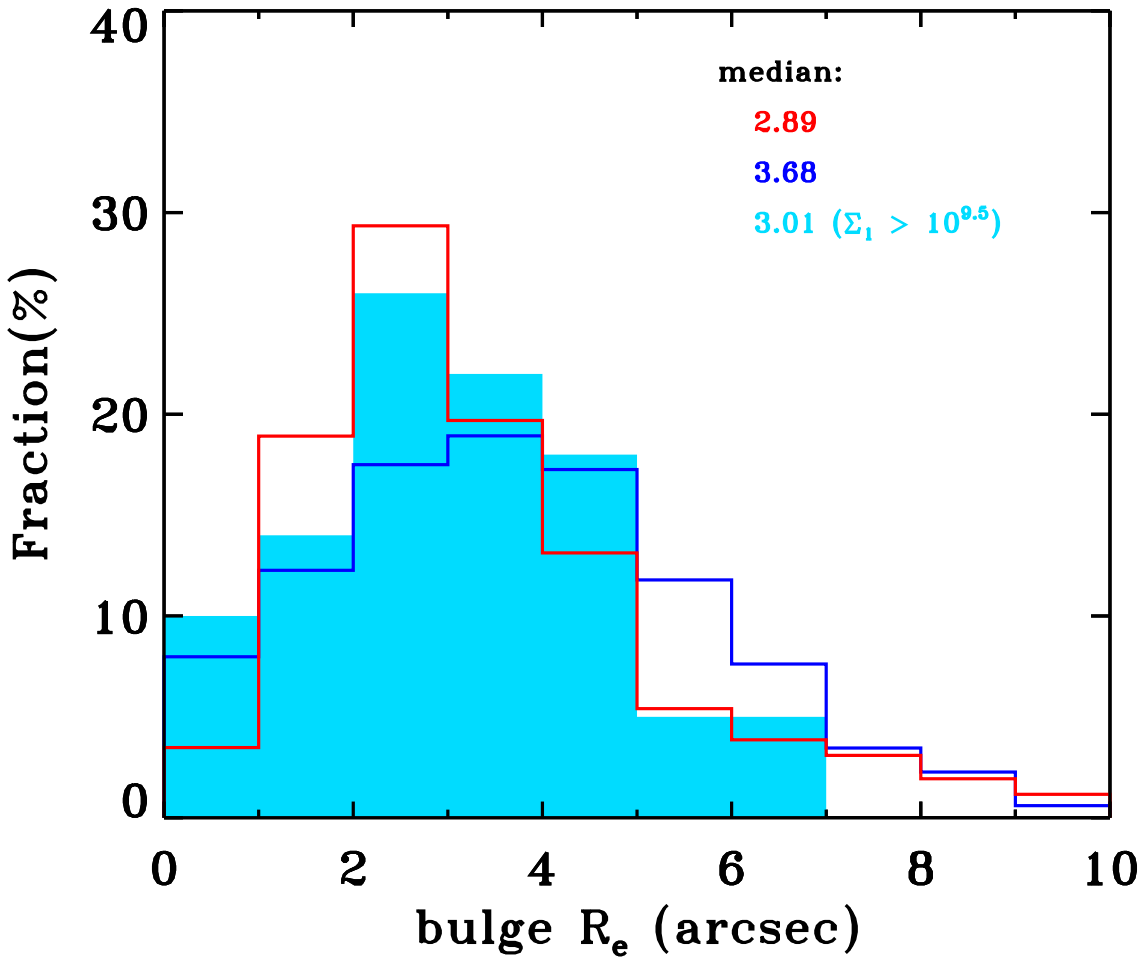}
\caption{Distributions of effective radius of bulges for our sample spiral galaxies.
The red and blue empty histograms represent spirals belonging to the
red sequence and blue cloud, respectively.
The light blue filled histogram represents blue spirals with
$\Sigma_1 > 10^{9.5} M_\odot \, {\rm kpc}^{-2}$.
The median values for the respective
categories are labeled at the top right corner.}
\label{bulgeRe.eps}
\end{figure}

The spectral indices D$_n(4000)$ and [Mg/Fe] measured from the central
3\arcsec\, diameter fiber spectrum were taken from the catalog of Max Planck
Instituted for Astrophysics-Johns Hopkins University
(MPA-JHU\footnote{http://www.mpa-garching.mpg.de/SDSS}) and the OSSY catalog,
respectively.  Specifically, we corrected the internal dust extinctions for
D$_n(4000)$ according to the Calzetti's law. The [Mg/Fe], defined as log
(Mgb/0.5(Fe5270 + Fe5335)) normalized to the solar abundances, were calculated
based on the Lick indices Mgb, Fe5270 and Fe5335 obtained from the OSSY
catalog, which improved on the MPA-JHU in the absorption line measurements,
especially in accounting for the impact of
[\ion{N}{1}]~$\lambda\lambda5198,5200$ lines on Mgb measurements. Since the
D$_n(4000)$ and [Mg/Fe] were measured from the central 3\arcsec\, diameter
fiber spectrum, they roughly represent the properties of the central regions of
galaxy bulges, as demonstrated by the bulge R$_e$
distribution for our red and blue spiral galaxies in Figure \ref{bulgeRe.eps}.
We note that the spectral indices measured from the central fiber spectrum may
suffer from contamination from the disk component, although we have attempted
to minimize this effect by excluding galaxies with $b/a < 0.5$.  However, this
effect should act on the blue and red spirals in the same way and hence does
not produce biased results in the comparisons.

The dark matter halo masses were extracted from the halo mass catalog of
\citet{Yang2007}, which were derived using statistical estimation for galaxies
in a group.  \citet{Yang2007} provides two sets of masses for each group based
on the characteristic stellar mass and the characteristic luminosity,
respectively. The dark matter halo mass based on the characteristic stellar
mass was adopted. There are 94.6\% (264/279) red spirals, 94.8\% (911/961)
blue spirals and 95.9\% (2,770/2,889) ellipticals in the group catalog,
among which 65.6\% (173/264) red spirals, 77.8\% (709/911) blue spirals and
68.1\% (1,886/2,770) ellipticals are central galaxies.
Only the dark matter halo masses of central galaxies will be explored.

The Arecibo Legacy Fast ALFA (ALFALFA) Survey provides a database of
extragalactic HI 21 cm line survey covering $\sim$ 7,000 deg$^2$
\citep{Giovanelli2005,Haynes2018}.  We cross-matched our samples with the
ALFALFA ($\alpha .100$) database using a 4$\arcsec$ searching radius to derive
the masses of the atomic HI gas. There are 59.5\% (166/279) red spirals, 58.3\%
(560/961) blue spirals and 56.3\% (1,627/2,889) elliptical galaxies in the
ALFALFA survey area, and 74 red spirals, 327 blue spirals and 39 ellipticals
have HI detections.

We measured the stellar mass surface density $\Sigma_1$ for our sample galaxies
using the five broadbands ($u,g,r,i,z$) SDSS Atlas images. We first retrieved
PSF Full Width at Half Maximum (FWHM) for the five bands from the "photoObjall"
catalog via the SDSS CasJobs service.  For each galaxy, we then performed PSF
matching between images in different bands by smoothing the four bands images
with better PSFs to the worst PSF using Gaussian kernels.  Then circular
aperture photometry with a radius of $1$\,kpc was carried out on the
psf-matched images.  The observed Spectral Energy Distribution (SED) consisting
of five broadband photometry was thus obtained and corrected for Galactic
reddening.  By comparing this SED to the SEDs in the model library using
Bayesian likelihood estimates \citep{Kauffmann2003b}, we derived the probability
distribution of the corresponding stellar mass within the central $1$\,kpc. We
used the median of the probability distribution as our best fit of the stellar
mass and the 16\% to 84\% values as the $\pm 1 \sigma$ errors. The details
about how to generate the model library can be found in \citet{Chen2012}.

\section{RESULTS\label{sec:result}}

In this work, we focus on investigating the formation and assembly processes as
well as the possible quenching mechanisms for massive red spiral galaxies, by
comparing the red spirals with blue spirals and ellipticals on their bulge
colors, $\alpha$-enhancement traced by [Mg/Fe] excess, age of stellar
populations indicated by D$_n(4000)$ and structures. We also concern the
similarities and differences among the red spirals, blue spirals and elliptical
galaxies, which might be able to provide some clues to understanding the whole
picture of massive galaxy formation and evolution.

\subsection{The Spectroscopic Properties \label{subsec:spectroproperties}}

\begin{figure}
\plotone{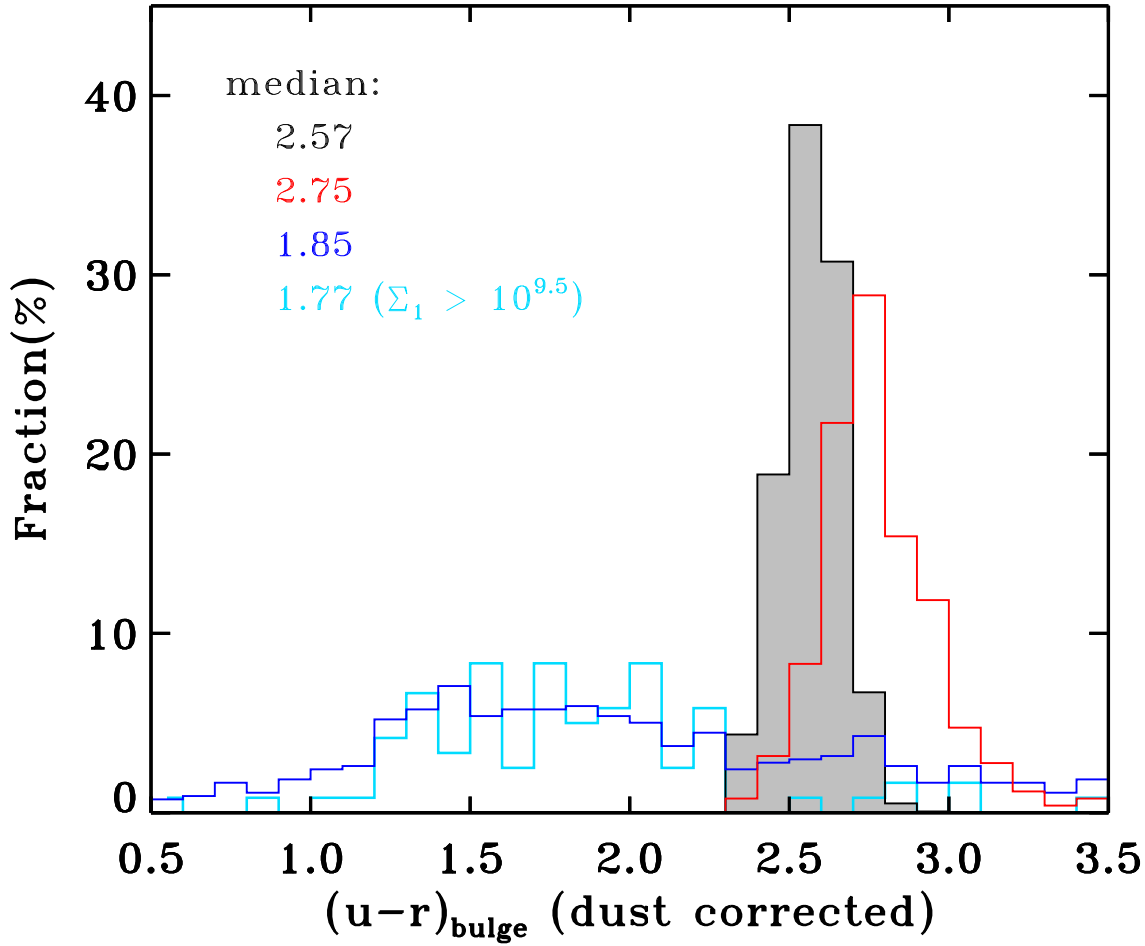}
\caption{Distributions of global {\em u-r} colors for the ellipticals (gray
filled histogram) and {\em u-r} colors of the bulges for red (red empty
histogram) and blue (blue empty histogram) spirals, as well as for the blue
spirals with $\Sigma_1 > 10^{9.5} M_\odot \, {\rm kpc}^{-2}$ (light blue empty
histogram).
The median colors are labeled at the top left corner.
The rms scatters of the color distributions for ellipticals, red, blue and high
$\Sigma_1$ blue spirals are 0.09, 0.14, 0.69 and 0.46, respectively.
The significance level is $> 99.99\%$ from the K-S test for the difference
between the color distributions of red and blue spirals.}
\label{urcordiff_Sbulge_redE_highmass.ps}
\end{figure}

Galaxy color is a direct probe of stellar populations.  Figure
\ref{urcordiff_Sbulge_redE_highmass.ps} shows the {\em u-r} color distributions
for the bulges of red and blue spirals, as well as the ellipticals. It is clear
from Figure \ref{urcordiff_Sbulge_redE_highmass.ps}
that the bulges of red spirals are even redder than ellipticals. The median
values of {\em u-r} colors for the bulges of red spirals and ellipticals are
2.75 and 2.57, respectively. Considering the negative color gradient of
early-type galaxies, we expect that the central regions of ellipticals are
redder than the entire galaxies and hence have similar {\em u-r} colors to the
bulges of red spirals.  Furthermore, both the bulges of red spirals and
ellipticals cover a relatively narrow range in {\em u-r} color with the rms
scatters of 0.14 and 0.09, respectively. In contrast, the color range for the
bulges of blue spirals is much larger with a rms scatter of 0.69, and the
median value of {\em u-r} color is 1.85 that is lower than those of red spirals and
ellipticals by $\sim33\%$ and $\sim28\%$, respectively. The
Kolmogorov-Smirnov (K-S) test shows that
the distributions of {\em u-r} color for red and blue spirals are completely
different with a significance of 99.99\%.  It implies that star formation is
still taking place in a large fraction of bulges of blue spirals, which is
consistent with their spectral features, such as strong H$\alpha$ emission
lines. 

\begin{figure}
\plotone{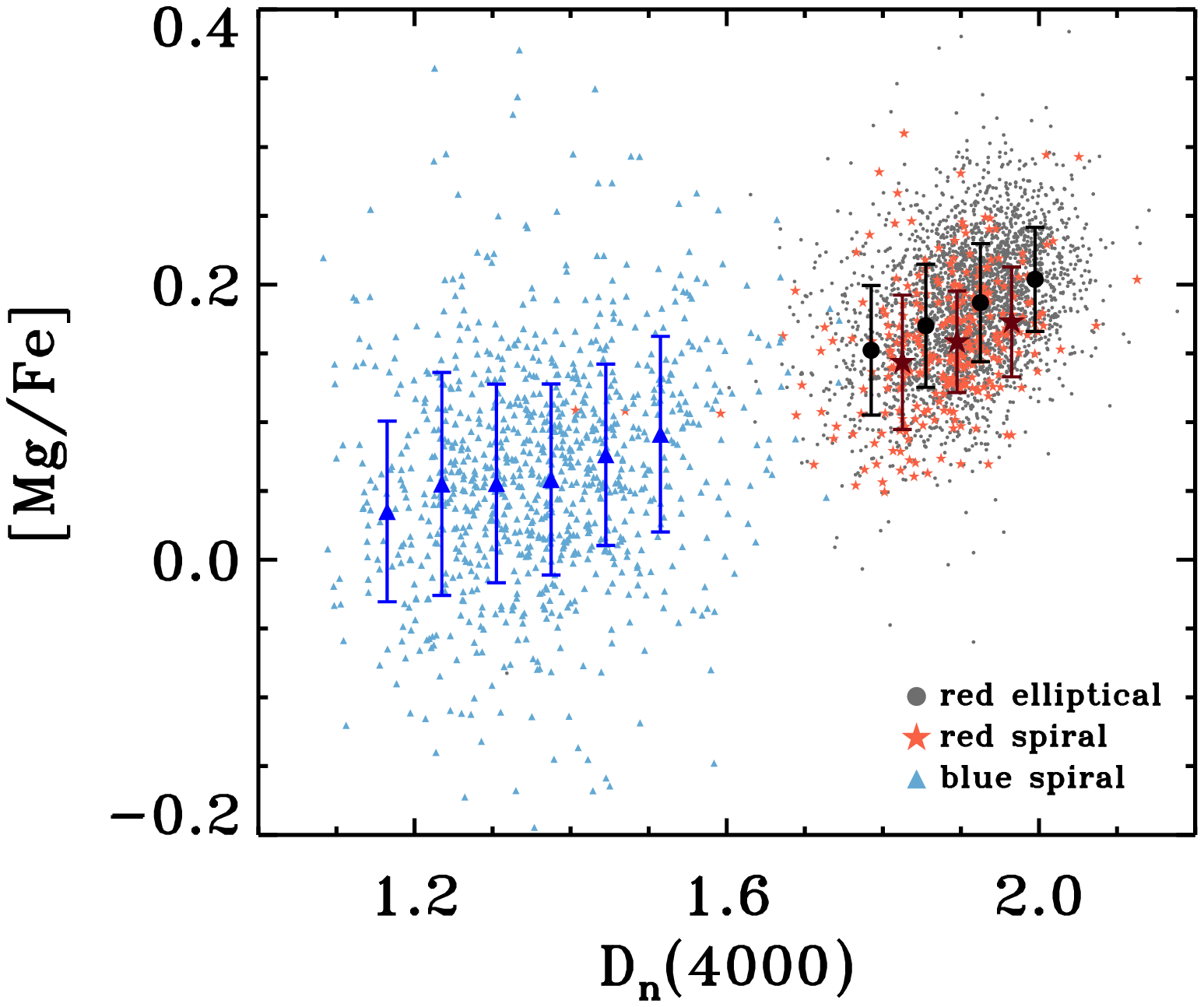}
\caption{[Mg/Fe] vs. D$_n(4000)$ relation for ellipticals (gray dots),
red (red stars) and blue (blue triangles) spirals.
Large black, dark red and dark blue symbols show the average values with
the rms scatters in each bin of D$_n(4000)$ for each sample.
The Spearman's rank order correlation coefficients for ellipticals, red and blue
spirals are 0.35, 0.32 and 0.21, respectively, with significance levels
$> 99.99\%$.}
\label{MgFeDn4000.ps}
\end{figure}

It is well known that $\alpha$-elements are mainly delivered by Type II
supernova explosions of massive stars, while a substantial fraction of Fe peak
elements comes from the delayed Type I supernova explosions. Therefore, the
$\alpha$/Fe ratio, indicated by [Mg/Fe], can reflect the relative importance of
Type II and Type I supernova in galaxies, and it carries the information of
star formation timescale in galaxies \citep{Thomas2005}.  On the other hand,
D$_n(4000)$ defined as the ratio of the average flux density in the bands
4000-4100\,\AA\ and 3850-3950\,\AA\ is a proxy for the galaxy age. Especially,
it is an excellent age indicator for old galaxies
\citep[e.g.,][]{Kauffmann2003b,Tacchella2017}.  Figure \ref{MgFeDn4000.ps} shows
the relation between [Mg/Fe] and D$_n(4000)$ for the red and blue spirals, as
well as ellipticals. 
%Given the large scatters in the distribution, we overlay the mean values and
%the rms scatters in bins of D$_n(4000)$ for the three types of galaxies to
%examine possible trends.  
We can see from Figure \ref{MgFeDn4000.ps} that the red spirals and ellipticals
occupy almost the same region in the [Mg/Fe] vs. D$_n(4000)$ diagram. Moreover,
both of them follow a trend that as D$_n(4000)$ increases, the [Mg/Fe] also
increases, although the correlations between [Mg/Fe] and D$_n(4000)$ are not
strong, with the Spearman's rank order correlation coefficients of 0.32 and
0.35, respectively. Since both D$_n(4000)$ and [Mg/Fe] were measured based on
the SDSS fiber spectra within a 3\arcsec {\,} diameter aperture, they probe the
stellar population properties of the central regions of our sample spirals and
ellipticals. Therefore, Figure \ref{MgFeDn4000.ps} indicates that the formation
epoch and formation duration for the main stellar populations in the central
regions of red spirals are similar to those of ellipticals.  Furthermore, the
correlations between [Mg/Fe] and D$_n(4000)$ for red spirals and ellipticals
illustrate that the earlier the main stellar population formed, the shorter the
star formation timescale was.  On the other hand, it is obvious from Figure
\ref{MgFeDn4000.ps} that the blue spirals are located in a completely different
region in the [Mg/Fe] vs.  D$_n(4000)$ diagram, and there is almost no
correlation between [Mg/Fe] and D$_n(4000)$, with the Spearman's rank
correlation coefficient of 0.21.

\begin{figure}
\plotone{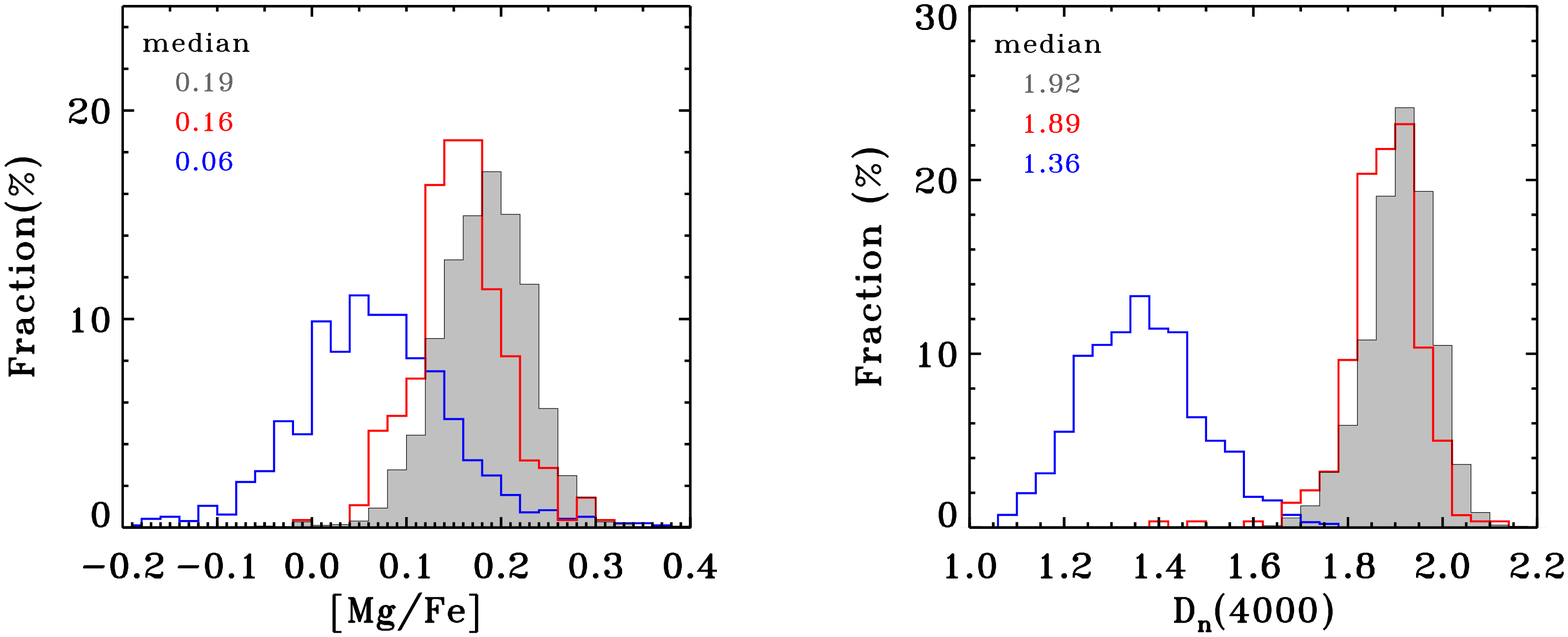}
\caption{Histograms of [Mg/Fe] ({\em left}) and D$_n(4000)$ ({\em right}) for
ellipticals, red and blue spirals. The colors of the histograms are the same as
in Figure \ref{newlgmdistri.ps}. The median values are shown at the
top left corner of each panel. The rms scatters of the distributions for
ellipticals in the [Mg/Fe] and D$_n(4000)$ are $\sim 0.05$ dex and
$\sim 0.07 \AA$, respectively.}
\label{MgFeDn4000_histo.ps}
\end{figure}

We also investigate the histograms of [Mg/Fe] and D$_n(4000)$ for
the bulges of red and blue spirals, as well as ellipticals, as shown in Figure
\ref{MgFeDn4000_histo.ps}. The median values of [Mg/Fe] and D$_n(4000)$ are
labeled in the figure, and the rms scatters of the distributions for
ellipticals in [Mg/Fe] and D$_n(4000)$ are $\sim$0.05 dex and $\sim$0.07
$\AA$, respectively. From the left panel of Figure \ref{MgFeDn4000_histo.ps},
the median values of [Mg/Fe] for the central regions of red spirals and
ellipticals are 0.16 and 0.19, respectively. They are in good agreement with
each other within the scatter.  In contrast, the median value of [Mg/Fe] for
the central regions of blue spirals is just 0.06, which is systematically
smaller than that of red spirals and ellipticals at $> 2\sigma$ levels.  As
\citet{Thomas2005} claimed, the longer the star formation timescale, the lower
is the [Mg/Fe] ratio. They also pointed out that $[{\rm Mg/Fe}]=0.2$
corresponds to a star formation timescale within $1$\,Gyr for composite stellar
populations.  Therefore, the star formation timescales for the central regions
of red spirals and ellipticals are similarly short, whereas the star formation
timescale for the bulges of blue spirals is longer. 

Furthermore, from the right panel of Figure \ref{MgFeDn4000_histo.ps}, the
median values of D$_n(4000)$ for the main stellar populations of the central
regions of red spirals and ellipticals are 1.89 and 1.92, respectively, which
are almost the same within the rms scatter of D$_n(4000)$.  This implies a
similar formation epoch of these two populations. In contrast, the median value
of D$_n(4000)$ for the main stellar populations in the central regions of blue
spirals is just 1.36 that is obviously smaller than that of red spirals and
ellipticals at $> 3\sigma$ levels. To avoid suffering from model dependence, we
adopted spectral indices to probe the properties of stellar populations in this
paper. From our experiment, the ages, especially the luminosity-weighted ages
derived from FIREFLY \citep{Comparat2017} produced consistent results with
D$_n(4000)$.

Given that the local massive ellipticals formed their main stellar populations
by redshift $\sim2$ and the star formation timescale is $\sim$ $1$\,Gyr
\citep[e.g.,][]{Worthey1992,Thomas2005}, the similar distributions of [Mg/Fe]
and D$_n(4000)$ for the central regions of red spirals and ellipticals strongly
suggest that the central regions of red spirals had also formed by redshift
$\sim2$ and within $\sim1$\,Gyr. These results support the conclusions for
bulge formation by \citet{Robaina2012}, \citet{Belli2015} and
\citet{Onodera2015}, also consistent with the recent analysis based on MaNGA
database \citep{Hao2019}.

However, the central regions of blue spirals formed later and the formation
timescale is longer than those of ellipticals as shown in Figure
\ref{MgFeDn4000.ps} and Figure \ref{MgFeDn4000_histo.ps}. Therefore, we need to
explore possible reasons that are responsible for the differences in stellar
populations of these different types of massive galaxies. It seems to be widely
accepted that galaxy quenching often associates with the construction of
central concentration, especially for massive galaxies \citep{Bell2012,
Cheung2012, Fang2013, Woo2015, Pan2016}.  Investigations for the central
structures and morphologies of massive spiral galaxies might help to gain
insights into the physical processes for galaxy formation.

\subsection{The Structure Properties of Massive Spiral Galaxies \label{subsec:structureproperties}}

\subsubsection{The Central Stellar Mass Density \label{subsubsec:density}}

It has been established that galaxy formation and quenching are closely related
to their central structures, represented by $\Sigma_1$, which is a key
parameter connecting the galaxy formation history
\citep[e.g.,][]{Fang2013,Barro2017a,Tadaki2017a,Whitaker2017,Tacchella2017, Chen2019}.  The
relation between $\Sigma_1$ and stellar mass has been widely investigated for
both local and high-redshift galaxies
\citep[e.g.,][]{Cheung2012,Fang2013,Barro2017a,Whitaker2017}.  It was found
that there is a tight correlation between $\Sigma_1$ and stellar mass for
quenched galaxies. The best-fitting line for quiescent galaxies is called the
$\Sigma_1$-$M_{\ast}$ ridgeline. In addition, star-forming galaxies tend to be
located below the $\Sigma_1$-$M_{\ast}$ ridgeline.

\begin{figure}
\plotone{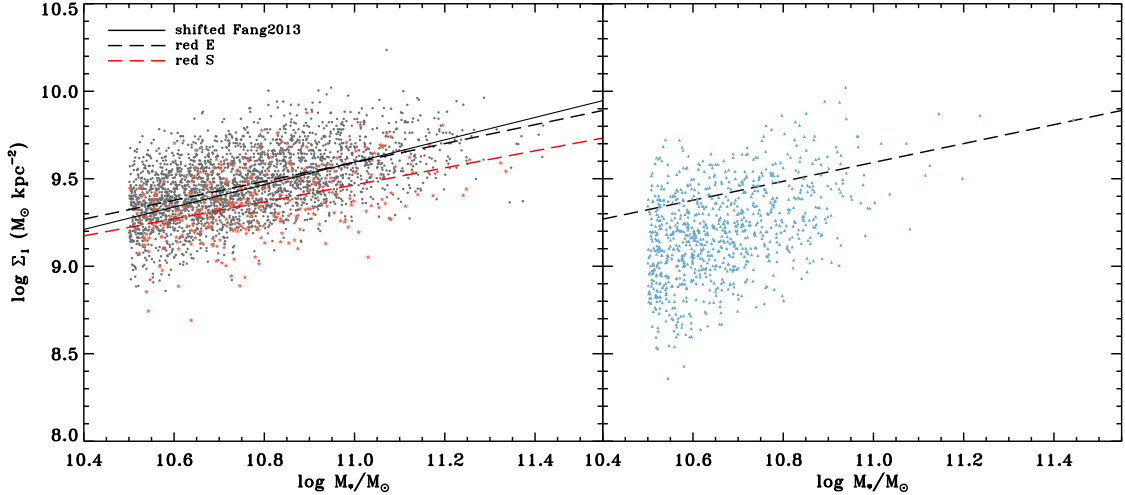}
\caption{
$\Sigma_1$ vs. stellar mass relation for red spiral and elliptical galaxies
({\em left}) and blue spiral galaxies ({\em right}). The symbols are
the same as in Figure \ref{MgFeDn4000.ps}.
The black solid line represents the $\Sigma_1$-$M_{\ast}$ ridgeline fitted by
Fang et al. (2013) but shifted by 0.15 dex and 0.04 dex on the vertical and
horizontal axes, respectively (see text). The red and black
dashed lines represent the ordinary least square fitting relations for red spirals
and ellipticals, respectively.
The 1$\sigma$ vertical scatters for red spirals and ellipticals are 0.17 and
0.16 dex, respectively.
The Spearman's rank correlation coefficients are 0.44 and 0.51 for red spirals
and ellipticals, respectively, with significance levels $> 99.99\%$.
}
\label{1kpc_totalmass_urred.ps}
\end{figure}

The left panel of Figure \ref{1kpc_totalmass_urred.ps} shows the $\Sigma_1$
versus $M_{\ast}$ relation for our sample massive red spirals and ellipticals.
It is clear that the red spirals are located similarly to the ellipticals.
Both of them show relatively strong correlations between $\Sigma_1$ and
$M_{\ast}$, with Spearman's rank correlation coefficients of 0.44 and 0.51,
respectively. For a direct comparison with \citet{Fang2013}, we adopt the
ordinary least square fitting method to derive the best-fit
$\Sigma_1$-$M_{\ast}$ relations and quote the associated vertical scatters
in this work. It turns out that the best ordinary least
square fitting lines for our red spirals and ellipticals are similar, with
consistent slopes within $1\sigma$ uncertainties and a slightly different
normalization in the sense that the normalization of the relation for red
spirals is 0.12 dex lower than the one for red ellipticals. We speculate
that the lower normalization for red spirals is caused by the presence of a
disk component in addition to the bulge component, and this will be tested
below. The best-fitted relation for ellipticals is in good agreement with that
of \citet{Fang2013} after being shifted by 0.15 dex and 0.04 dex on the
vertical and horizontal axes, respectively. Such offsets in the
$\Sigma_1$-$M_{\ast}$ relation are caused by the different methods adopted in
the stellar mass measurements. In addition, the $1$\,$\sigma$ vertical
scatter about the relation for our ellipticals is exactly the same as that of
\citet{Fang2013}, with a value of 0.16 dex, although the sample selection
criteria of these two studies are different. Interestingly, the red spirals
show a very similar vertical scatter of 0.17 dex. In contrast, the blue spirals
are located systematically lower than the $\Sigma_1$-$M_{\ast}$ ridgeline and
show a larger scatter, as shown in the right panel of Figure
\ref{1kpc_totalmass_urred.ps}.

\begin{figure}
\plotone{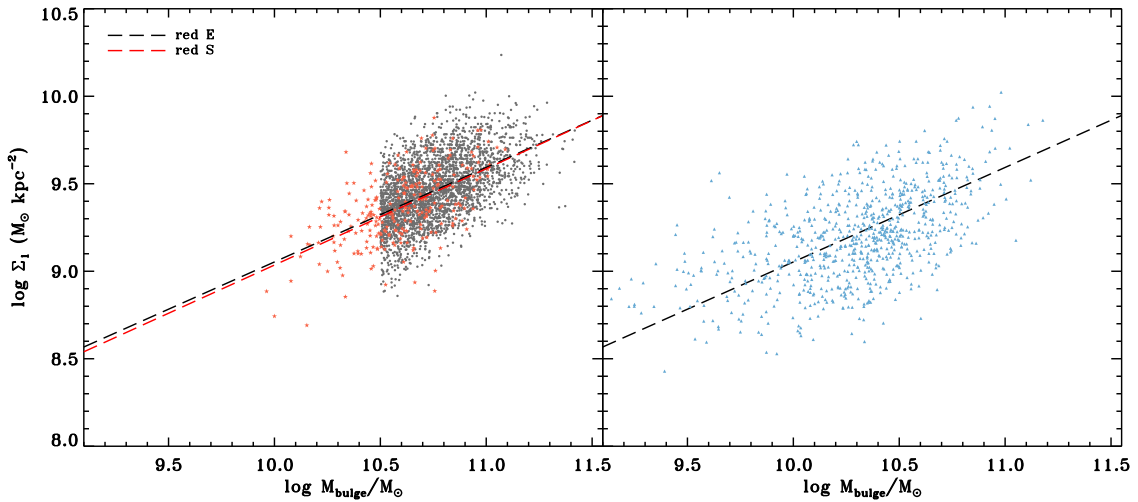}
\caption{
$\Sigma_1$ vs. bulge mass relation for red spiral and elliptical galaxies
({\em left}) and blue spiral galaxies ({\em right}). The symbols are
the same as in Figure \ref{MgFeDn4000.ps}. The red and black
dashed lines represent the ordinary least square fitting relation based on red spirals and
ellipticals, respectively.
The 1$\sigma$ vertical scatters for ellipticals, red and blue spirals
are 0.16, 0.16 and 0.21 dex, respectively.
The Spearman's rank correlation coefficients are 0.51, 0.54 and 0.53 for
ellipticals, red and blue spirals,
respectively, with significance levels $> 99.99\%$.
}
\label{1kpc_bulgemass_urred.ps}
\end{figure}

It is very interesting that if we just concern the bulge mass of red spirals,
we can see from the left panel of Figure \ref{1kpc_bulgemass_urred.ps} that the
bulges of red spirals and ellipticals follow the same $\Sigma_1$-$M_{\ast}$
relation in terms of both the slope and the intercept. Their Spearman's rank
correlation coefficients are also very similar, i.e., 0.54 and 0.51,
respectively, with 1$\sigma$ vertical scatters of 0.16 dex. It indicates that
the bulges of red spirals share the $\Sigma_1$-$M_{\ast}$ relation with
quenched galaxies.  Note that all the above correlation analyses show
significance levels larger than 99.99\%.  Altogether, it illustrates that the
central structure of red spirals is more closely connected with the bulge mass,
instead of the mass of the whole galaxy. This confirms our above
speculation that the disk component is responsible for the relatively lower
normalization of the $\Sigma_1$-$M_{\ast}$ relation for red spirals compared to
that of the ellipticals. This can be easily understood in consideration of the
results in Section 3.1.  The central parts of bulges of red spirals formed
rapidly by $z\sim2$, then the disks of the galaxies started to grow also fast
by gas falling from larger radii during very gas-rich mergers \citep{Hao2019}.  

\begin{figure}
\plotone{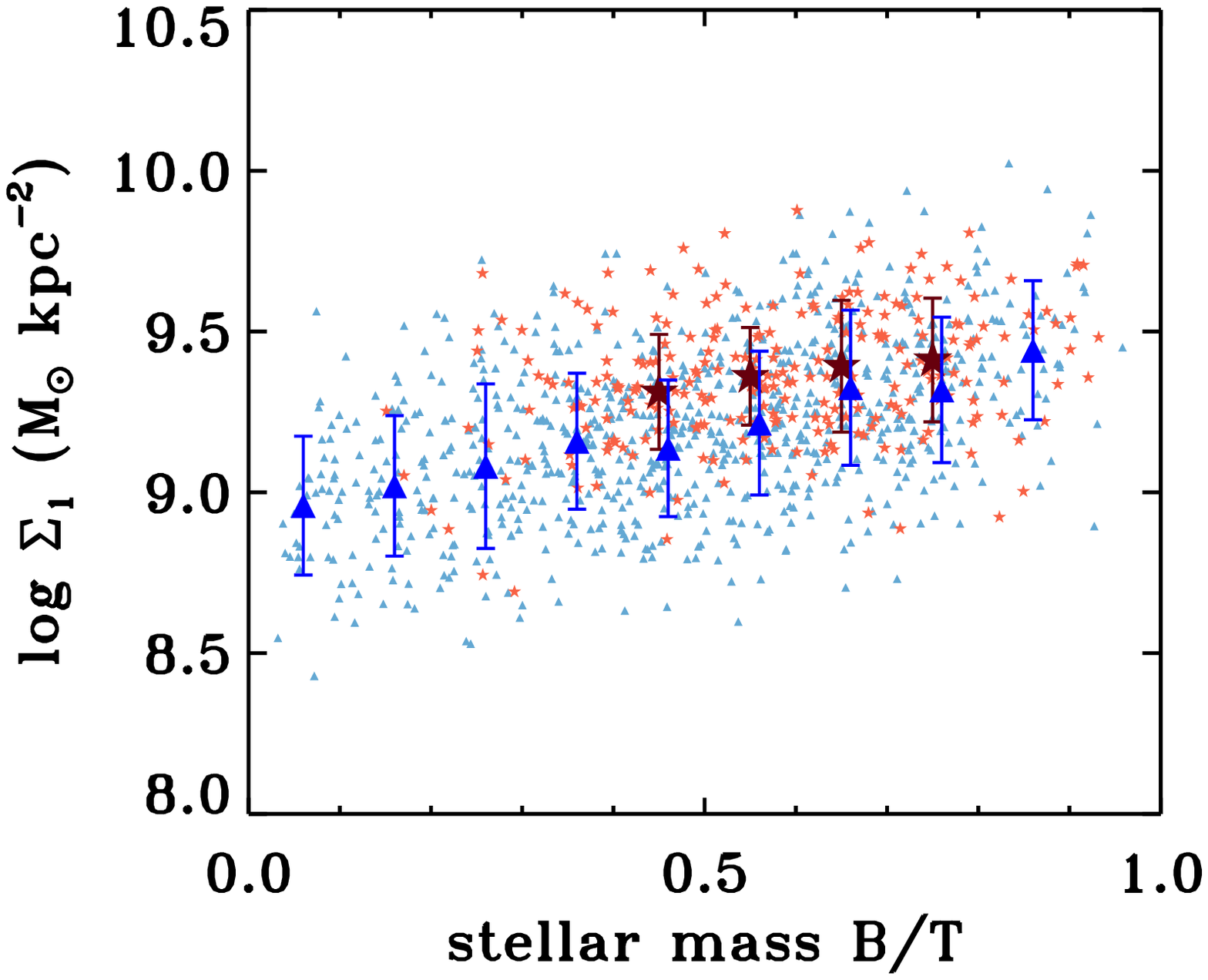}
\caption{
$\Sigma_1$ vs. B/T relation for red and blue spiral galaxies.
The symbols are the same as in Figure \ref{MgFeDn4000.ps}.
The Spearman's rank correlation coefficients are 0.26 and 0.45 for red and blue
spirals, respectively, with significance levels $> 99.99\%$.}
\label{Sigma1_BTmassBD.eps}
\end{figure}

Moreover, we examine the $\Sigma_1$-$M_{\ast}$ relation for the bulges of blue
spirals, as shown in the right panel of Figure \ref{1kpc_bulgemass_urred.ps}.
It can be seen that there is also a correlation between $\Sigma_1$ and
$M_{\ast}$ for the bulges of blue spirals, with the Spearman's rank correlation
coefficient of 0.53 at a significance level $> 99.99\%$.  The 1$\sigma$ vertical
scatter for blue spirals is 0.21 dex, which is larger than that of red spirals
by $\sim 30\%$.  Taken together, the bulges of both massive red and blue
spirals follow the $\Sigma_1$-$M_{\ast}$ ridgeline for quenched galaxies. The
behavior of the bulges of blue spirals is distinct from that of the entire blue
spiral galaxies, which are mainly located below the $\Sigma_1$-$M_{\ast}$
ridgeline for quenched galaxies, as shown in the right panel of Figure
\ref{1kpc_totalmass_urred.ps}. The close correlation between $\Sigma_1$ and the
bulge mass for both red and blue spirals may support the statement of
\citet{Bluck2014} that bulge mass is the king for galaxy quenching. The star
formation in high $\Sigma_1$ blue spirals is very likely due to rejuvenation. We will
discuss this possibility in Section 3.2.3.

As another popular probe of galaxy bulges, $B/T$ is also expected to be
correlated with $\Sigma_1$.  Figure \ref{Sigma1_BTmassBD.eps} shows the
$\Sigma_1$-$B/T$ relation for the red and blue spirals. There is a weak
correlation between $\Sigma_1$ and $B/T$ for red spirals and a relatively
strong correlation for blue spirals, with the Spearman's rank correlation
coefficients of 0.26 and 0.45, respectively. Therefore, the central structure
represented by $\Sigma_1$ is more closely coupled with the bulge mass than the
bulge to total mass ratio $B/T$, which reinforces the importance of bulge mass
once more.

\subsubsection{The Morphologies of Massive Red Spiral Galaxies \label{subsec:morphology}}

More and more pieces of evidence suggest that massive elliptical galaxies
formed mainly via two phases
\citep[e.g.,][]{Oser2010,vanDokkum2010,HuangSong2018,Tanaka2019,Zibetti2020}.
During the first phase, the central compact region formed rapidly by cold gas
falling triggered by violent disk instabilities or gas-rich mergers
\citep{DekelBurkert2014,Zolotov2015} by redshift $\sim2$. In the second phase,
the outer extended part emerged gradually by accreting surrounding gas-poor
satellites or minor dry mergers \citep[e.g.,][]{Naab2009}. On the other hand,
the formation process of massive spiral galaxies, especially massive red spiral
galaxies, is still at debate \citep[e.g.,][]{Bundy2010,Masters2010,Hao2019}.
The disk formation models have been proposed since 1980's
\citep{Fall1980,Mo1998,Dutton2007}. It was claimed that galactic disks formed
from the dissipational collapse of gas in dark-matter halos and then evolved
secularly with quiet merging histories. More recently, numerical simulations
showed another possibility, in which very gas-rich major mergers can also
produce spiral galaxies \citep{Springel05, Robertson06, Hopkins09,
Athanassoula2016, SparreSpringel2017}. In this scenario, the bulge formed first
by cold gas falling into the center with a starburst mode, and then the gas at
sufficiently large radii cool quickly and re-form a rotating disk.  Such
merging events should have imprinted on the morphologies of galaxies.
Therefore, we study the morphologies and detailed structures of red spirals to
understand the possible mass assembly processes for their disks and outer
parts.

We visually inspect the SDSS images of our 279 sample red spirals very
carefully, also consulting the classification results by GZ 2
\citep{Willett2013, Hart2016}.  The morphologies of red spirals can be roughly
classified into three categories: 1) Galaxies with strong bars, inner and outer
rings (or shells); 2) Interacting galaxies/mergers; 3) Normal spiral galaxies
with bulge, disk and clear spirals arms.  For the reliability of the
classification of galaxies in the second category, we checked the redshifts of
their neighbors.  When the velocity difference between the target galaxy and
its neighbor is less than 500 km/s, the target galaxy is classified as an
interacting galaxy/merger.  For galaxies with a target-neighbor velocity
difference larger than 500 $km/s$ or without redshift information available for
their neighbors, only those with clear merging features, such as tidal tails,
are classified as mergers.

\begin{figure}
\plotone{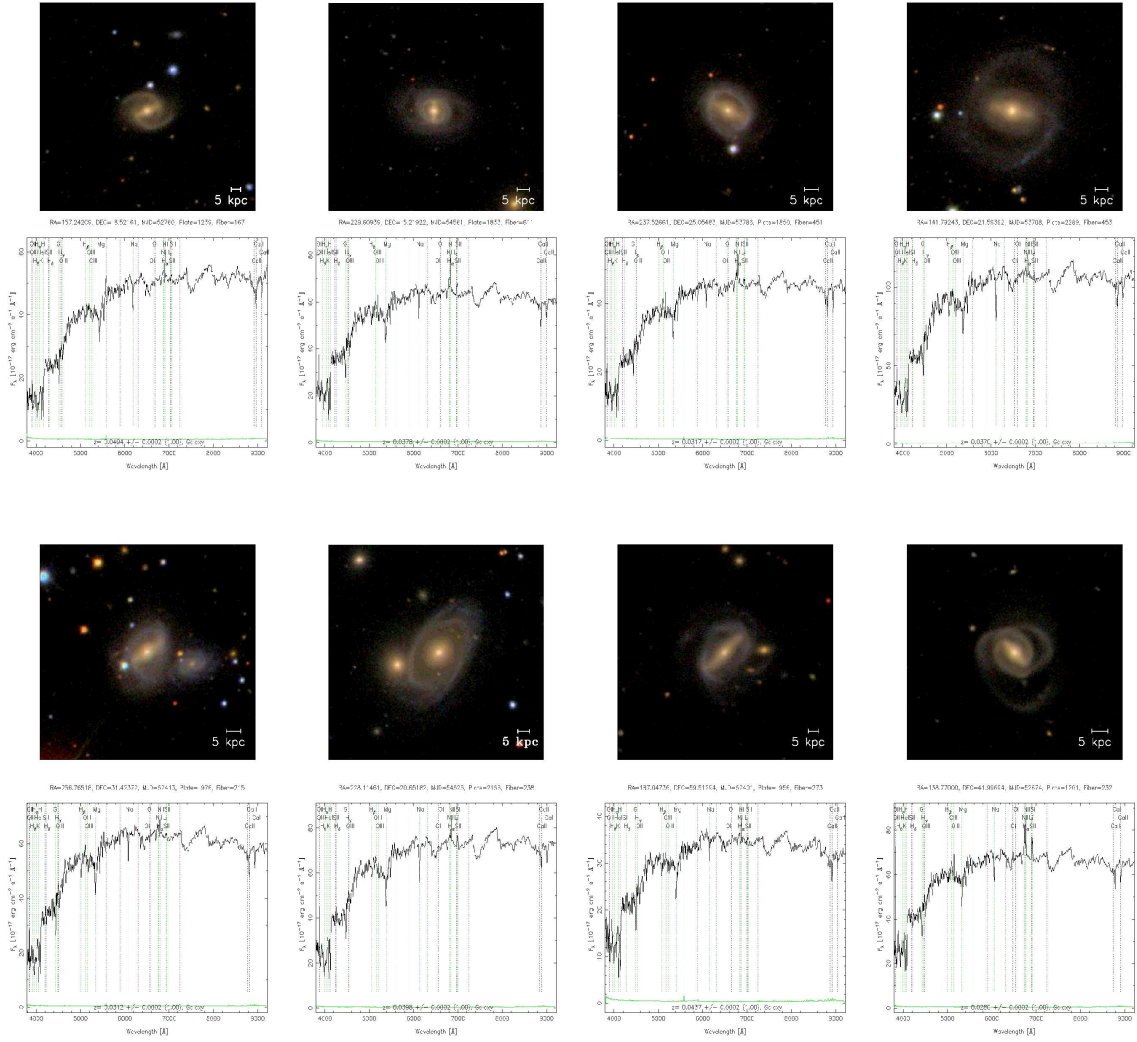}
\caption{Example true color SDSS images and spectra of massive spirals in the
red sequence. The top row shows galaxies with strong bars, inner and outer
rings or shells. The bottom row shows galaxies with
interacting or merging features. The scale bar at the bottom right
corner of each panel indicates a physical scale of 5\,kpc.}
\label{imgspec-red.ps}
\end{figure}

The fractions of these three categories are about 50\%, 20\% and 30\%,
respectively.  The top row of Figure \ref{imgspec-red.ps} shows the first class
of our sample red spirals, i.e., the galaxies with strong bars, inner and outer
rings (or shells), while the bottom row gives examples for the interacting or
merging galaxies.  It is obvious from the top row of Figure
\ref{imgspec-red.ps} that the rings, especially the outer rings (or shells) are
symmetric. In contrast, the galaxies in the bottom row show asymmetric or
incomplete outer rings, which are the tidal streams. It seems that these tidal
streams are in the process of forming rings/shells.  By comparing the
morphologies of red spirals in the first and second categories, we speculate
that rings or shells might be produced in mergers.

Recent deep observations have revealed that tidal streams and shells around
massive early-type galaxies are popular
\citep[e.g.,][]{Tal2009,Atkinson2013,Hood2018}.  On the other hand, numerical
simulations, especially, the recent cosmological hydrodynamical zoom-in
simulations, such as EAGLE, Illustris, Horizon-AGN, revealed the role of galaxy
interaction and merger on shaping galaxy morphologies. Their results showed
that the vast majority of ring galaxies formed via interactions of galaxies
\citep{Elagali2018}, the most strong bars in the local universe are triggered
by galaxy mergers or external perturbations \citep{PeschkenLokas2019}, and the
gas-rich major mergers can form rotationally supported structure by re-growing
a disk
\citep[e.g.,][]{SparreSpringel2017,Athanassoula2016,Rodriguez-Gomez2017,Martin2018}.
Therefore, our morphological results are consistent with the simulations, i.e.,
interactions and mergers probably play an important role in the formation of
the disk and outer parts of red spirals. 

Furthermore, we can also see from Figure \ref{imgspec-red.ps} that the massive
red spiral galaxies of all kinds of morphologies have typical spectra of old
stellar populations with very weak or no emission lines, consistent with the
bulge $u-r$ color distributions. Both indicate that the bulges of massive red
spiral galaxies have been quenched. 

\begin{figure}
\plotone{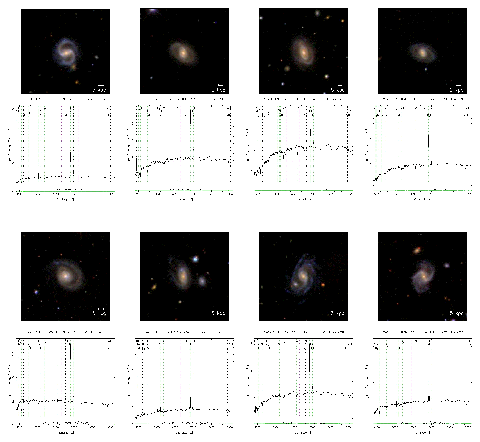}
\caption{Example true color SDSS images and spectra of massive blue spirals
with $\Sigma_1 > 10^{9.5} \, M_\odot \, {\rm kpc}^{-2}$.
The top row shows galaxies with strong bars, inner and outer rings or shells.
The bottom row shows galaxies with
interacting or merging features. The scale bar at the bottom right
corner of each panel indicates a physical scale of 5\,kpc.}
\label{imgspec-blue1.ps}
\end{figure}

\subsubsection{The High $\Sigma_1$ Massive Blue Spiral Galaxies \label{subsec:blue}}

As shown in the right panel of Figure \ref{1kpc_totalmass_urred.ps}, most blue
spirals are located below the $\Sigma_1$-$M_{\ast}$ ridgeline.  However, there
do exist some high $\Sigma_1$ blue spirals lying on the $\Sigma_1$-$M_{\ast}$
ridgeline.  It means that these blue spirals have already acquired dense cores,
but have not been quenched.  \citet{Fang2013} already noted this and speculated
that high $\Sigma_1$ blue spirals may be rejuvenated galaxies from red to blue
at fixed stellar mass but without central mass growth. However, more pieces of
evidence are needed.  We investigate the high $\Sigma_1$ (defined as galaxies
with $\Sigma_1$ larger than $10^{9.5} M_\odot \, {\rm kpc}^{-2}$) blue spirals by
comparing their morphology, structures traced by bulge mass and bulge-to-total
mass ratio ($B/T$), as well as the dark matter halo mass to those of red
spirals and ellipticals.

The high $\Sigma_1$ blue spirals comprises $\sim$ 10\% of our sample of blue
spirals. We classified their morphologies into three categories as for the red
spirals in the above subsection.  Figure \ref{imgspec-blue1.ps} shows the
examples of the first and second categories, i.e., the galaxies with strong
bars, inner and outer rings (top row), and those with interacting or merging
features (bottom row). The fractions of the three categories are almost the
same as those of red spirals, i.e., $\sim 2/3$ of high $\Sigma_1$ blue spirals
are with strong bars, inner and outer rings or in the interacting/merger stage.
However, when we perform the same classification for our sample blue spirals
with $\Sigma_1$ less than $10^{9.5} M_\odot \, {\rm kpc}^{-2}$ (denoted as low
$\Sigma_1$, hereafter), the fractions belonging to the three categories are
about 20\%, 15\% and 65\%, respectively. The fact that most low $\Sigma_1$ blue
spirals are normal spiral galaxies is significantly different from that of the
high $\Sigma_1$ blue spirals. 

\begin{figure}
\plotone{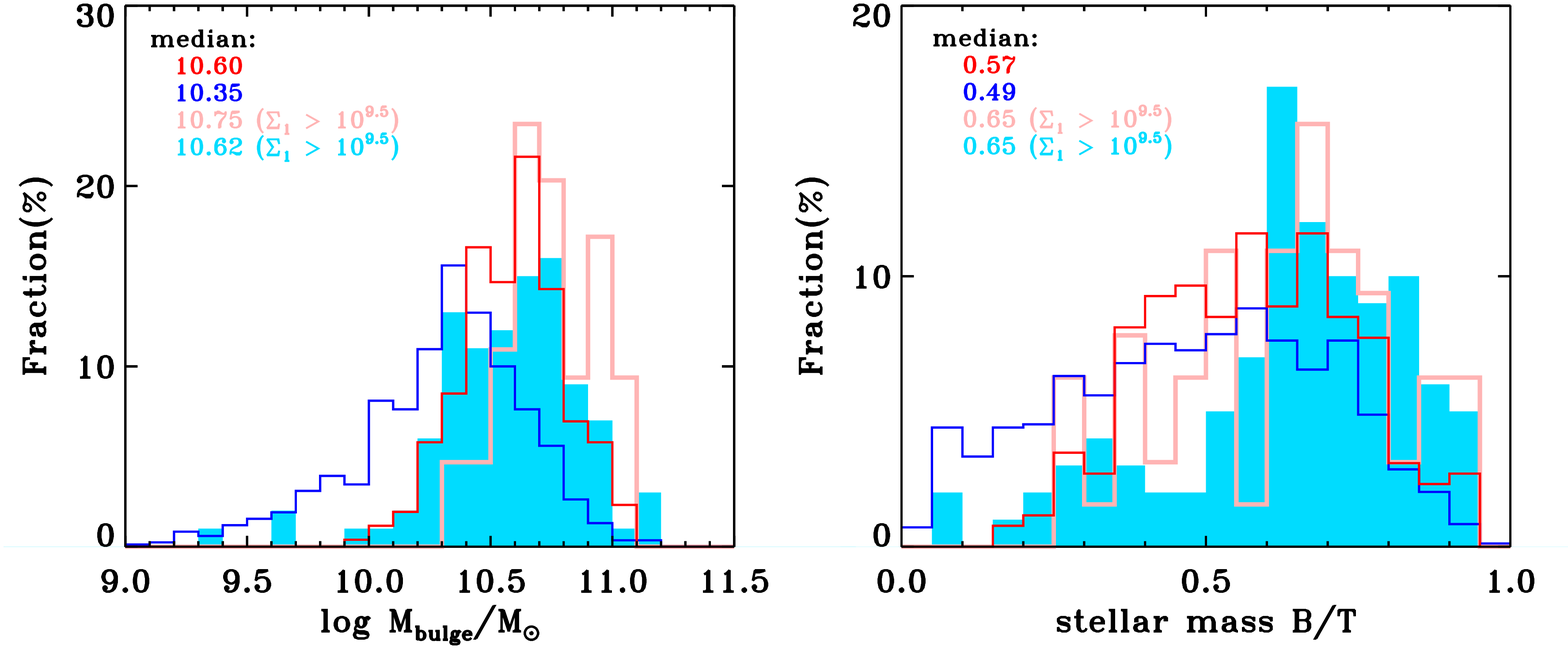}
\caption{Bulge mass (left) and stellar mass B/T (right) distributions for red
(red empty histogram) and blue (blue empty histogram) spirals,
as well as for the red (pink empty histogram) and blue (light blue filled
histogram) spirals with $\Sigma_1 > 10^{9.5} \, M_\odot \, {\rm kpc}^{-2}$.
The median values are labeled at the top left corner of each panel.
The rms scatters of the bulge mass distributions for red and blue spirals,
high $\Sigma_1$ red and blue spirals are 0.21, 0.33, 0.18
and 0.24 dex, respectively.
The rms scatters of the stellar mass B/T distributions for the galaxies belonging
to these four categories are 0.17, 0.22, 0.18 and 0.20 dex, respectively.
}
\label{BThigh.ps}
\end{figure}

Figure \ref{BThigh.ps} shows the distributions of the bulge mass (left panel)
and the bulge to total stellar mass ratio $B/T$ (right panel) for red and blue
spirals, as well as for the red and blue spirals with high $\Sigma_1$.  From
the left panel of Figure \ref{BThigh.ps}, the rms scatters of the bulge mass
distributions for the red spirals and the high $\Sigma_1$ red and blue spirals
are similar, i.e., $\sim 0.2$ dex, but the rms scatter for the blue spirals is
obviously larger, with a value of 0.33 dex.  The median values of the
distributions, as shown in the left panel of Figure \ref{BThigh.ps}, tells that
the bulge mass of the red spirals ($10^{10.60}M_{\odot}$) is larger than that
of the blue spirals ($10^{10.35}M_{\odot}$) by 0.25 dex, slightly larger than $
1 \sigma$.  In comparison, the median value of the bulge mass for high
$\Sigma_1$ blue spirals ($10^{10.62}M_{\odot}$) is almost the same as that of
the red spirals, although it is smaller than that of the high $\Sigma_1$ red
spirals by 0.13 dex.  From the right panel of Figure \ref{BThigh.ps}, it is
interesting to note that the median values of the $B/T$ ratios for high
$\Sigma_1$ red and blue spirals are the same (0.65), and even a little higher
than that of the whole population of red spirals (0.57). Furthermore, the
fractions of high $\Sigma_1$ red and blue spirals with $B/T > 0.6$ are larger
than those of the whole population of red and blue spirals.  In contrast, blue
spirals shows smaller median value of the $B/T$ ratio than all the other three
subsamples, and have the smallest fraction of galaxies with $B/T > 0.6$.
Therefore, there is no doubt that bulges are the dominant component and
contributes $\sim 2/3$ of the total stellar mass for both high $\Sigma_1$ red
and blue spirals.  Although blue spirals also harbor massive bulges, their
bulges are less prominent and less massive than those of high $\Sigma_1$ red
and blue spirals.

\begin{figure}
\plotone{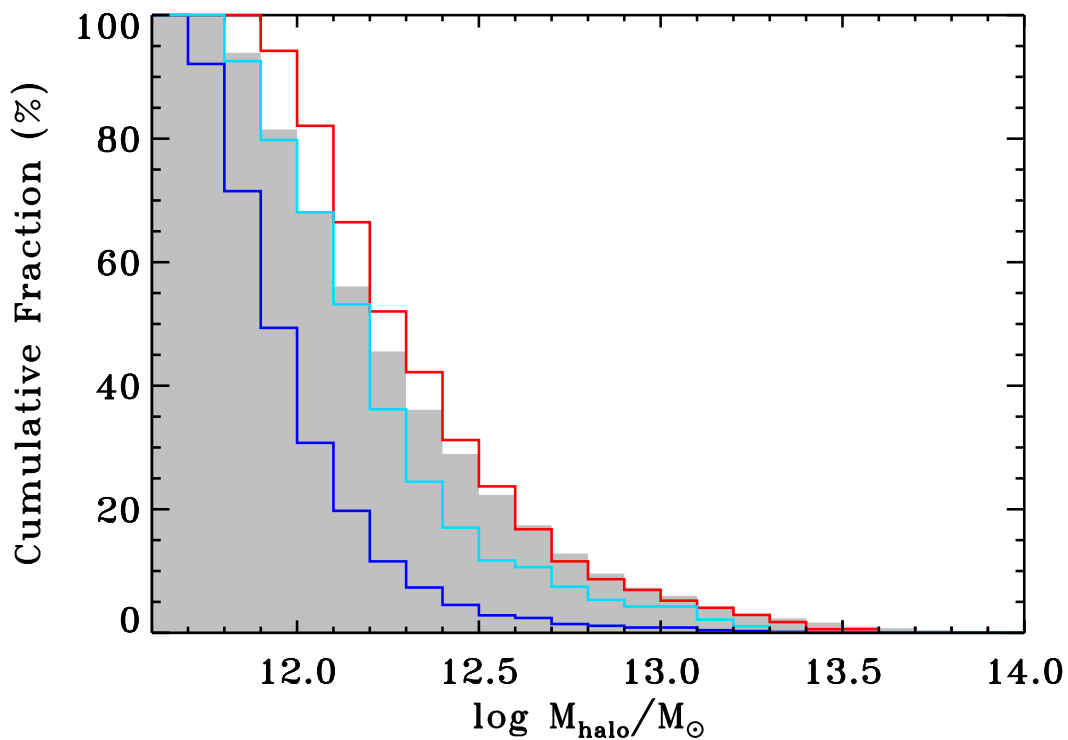}
\caption{Cumulative fractions of halo mass for the central galaxies in our
sample red (red line) and blue (blue line) spiral galaxies,
elliptical galaxies (gray filled histogram) and
blue spirals with $\Sigma_1 > 10^{9.5} \, M_\odot \, {\rm kpc}^{-2}$
(light blue empty histogram).}
\label{halo.ps}
\end{figure}

Figure \ref{halo.ps} shows the cumulative fractions of the halo mass of the
central galaxies for our sample red and blue spirals, as well as the elliptical
galaxies. We can see that the halo masses of the red spirals and ellipticals
are in a similar range, but are significantly larger than those of the blue
spirals. More than 80\% of the red spirals have halo masses larger than the
critical halo mass of $10^{12} M_{\odot}$, whereas such fraction reduces to
30\% for the blue spirals. It is especially intriguing when we examine the
cumulative distribution of the halo mass for the high $\Sigma_1$ blue spirals.
The high $\Sigma_1$ blue spirals show a similar distribution to that of red
spirals at $M_{\rm halo} < 10^{12.2} M_{\odot}$, but their halo masses are
significantly larger than the whole population of the blue spirals.

Given the blue color of the high $\Sigma_1$ blue spirals, there must be
star formation taking place in them. Figure \ref{imgspec-blue1.ps} also shows
that the high $\Sigma_1$ blue spirals possess blue disks, shells and spiral
arms, indicating on-going star formation activities. We then further examine
the central fiber spectra of high $\Sigma_1$ blue spirals. From the examples
shown in Figure \ref{imgspec-blue1.ps}, it is obvious that there are strong
Balmer emission lines in the spectra. Considering that the fiber spectra may
also include emission from the bar or disk, we examine
the bulge R$_e$ of the high $\Sigma_1$ blue spirals, as shown in Figure
\ref{bulgeRe.eps}. The median value of R$_e$ for the high $\Sigma_1$ blue
spirals is about 3\arcsec\ , which is twice of the fiber radius, and thus the
contamination from disk emission should not be significant. The presence of
star formation in bulges is further demonstrated by the bulge color
distribution of the high $\Sigma_1$ blue spirals in Figure
\ref{urcordiff_Sbulge_redE_highmass.ps}. From Figure
\ref{urcordiff_Sbulge_redE_highmass.ps}, we can see that the vast majority of
the bulges of high $\Sigma_1$ blue spirals are bluer than the bulges of red
spirals and have comparably blue {\em u-r} colors to those of blue
spirals. These results are in stark contrast to red spirals whose bulges have
been fully quenched.

In summary, there are similarities between the high $\Sigma_1$ blue and red
spirals in morphology, bulge mass and $B/T$ distributions, and halo mass range,
which strongly suggest that there are physical connections between these two
populations. The only difference between them is in the star formation
properties. Compared to the red spirals whose star formation has been quenched,
the high $\Sigma_1$ blue spirals host on-going star formation throughout the
galaxy, even within their bulges, as exhibited by their optical spectra and
{\em u-r} colors.

On the other hand, the central velocity dispersion was found to be tightly
correlated with $\Sigma_1$ for red sequence galaxies, with a rms scatter of
0.18 dex in  $\Sigma_1$ \citep{Fang2013}.  In fact, the central velocity
dispersion $\sigma_0$ has long been used to study its correlations with stellar
population properties \citep[e.g.,][]{Burstein1988,Bender1993}. For example,
\citet{Thomas2005} pointed out that at a given $\sigma_0$, the stellar
populations of the bulges of spirals are indistinguishable from those of the
ellipticals.  \citet{Wake2012} claimed that $\sigma_0$ is the best indicator of
galaxy color and is correlated with halo and central black hole mass.  Most
recently, \citet{Bluck2019} emphasized that $\sigma_0$ is the best parameter to
predict quenching for central galaxies. These results and the tight correlation
between $\sigma_0$ and $\Sigma_1$ strengthen that both $\sigma_0$ and
$\Sigma_1$ are indicators of galaxy quenching. It is just because $\Sigma_1$ is
easier to be measured than $\sigma_0$ that $\Sigma_1$ is used more widely as a
quenching indicator nowadays. The large $\Sigma_1$ values of the high
$\Sigma_1$ blue spirals possibly suggest that they were ever quenched before
the re-ignition of the star formation.  The similarities of the high $\Sigma_1$
blue and red spirals in morphology, structure and halo mass range provide a
further suggestion that the ever quenched high $\Sigma_1$ blue spirals were red
spirals indeed. Therefore, one of the reasonable interpretations for the
similarities between the high $\Sigma_1$ blue and red spirals is that high
$\Sigma_1$ blue spirals are rejuvenated red spirals, as induced by bar
instabilities or interactions, which lead cold gas to fall into the galaxies
and trigger star formation again in the bulges and disks of red spirals,
driving them towards blue. However, further careful investigations on the star
formation histories of high $\Sigma_1$ blue spirals are needed to validate this
scenario, as those did in \citet{Mancini2019} or in \citet{Hao2019}.

\section{DISCUSSION\label{subsec:ellipticals}}

Our results showed that red spirals and ellipticals have similar integrated
$u-r$ colors, central stellar population properties and a dark matter halo mass
range, and harbor similarly compact cores with high stellar mass surface
densities measured by $\Sigma_1$, but their outer structures are completely
different, the rotational disks hosted by red spirals in particular. This may
hint that there are both similarities and differences in the formation and
quenching processes for massive ellipticals and spirals. An investigation into
their gas properties may help us shed light on this issue.

\begin{figure}
\plotone{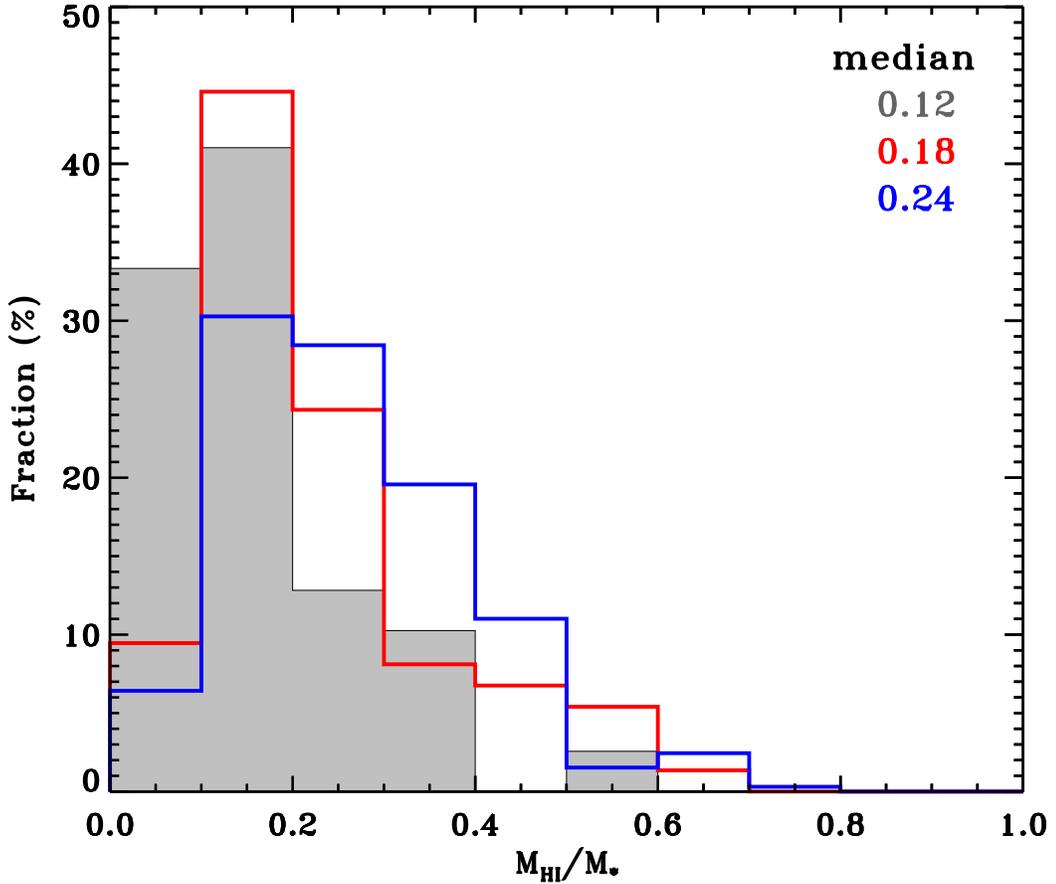}
\caption{HI gas mass to stellar mass ratio for HI detected sample galaxies.
The colors of the histograms are the same as in Figure \ref{newlgmdistri.ps}.
The median values for the respective categories are labeled at the top right
corner.}
\label{fHI_hist.eps}
\end{figure}

We investigate the atomic gas content in our sample galaxies based on the
ALFALFA ($\alpha .100$) data, which has a beam size of $3\farcm 8 \times
3\farcm 3$. As expected, the HI detection rate for ellipticals is very low
(2\%). However, it is surprising that the HI detection rate for our red spirals
is not too different from that of the blue spirals, i.e., 45\% and 58\%,
respectively, considering that red spirals have been quenched. A further
examination on the distribution of gas to stellar mass ratio $M_{HI}/M_{\ast}$
for all types of HI-detected sample galaxies was shown in Figure
\ref{fHI_hist.eps}. The median HI mass fraction for the 2\% HI-detected
ellipticals is only about 10\%, while the median HI mass fractions for all
types of HI-detected spiral galaxies are about 20\%, which is consistent with
the earlier works \citep[e.g.,][]{Gereb2018,Parkash2019,Zhang2019}. The similar
HI detection rate and HI mass to stellar mass ratio of red and blue spirals
seem to conflict with the halo quenching scenario, considering that red spirals
are mostly hosted by dark matter halos more massive than the critical mass
($\sim 10^{12}\,M_\odot$) for quenching, as shown in Figure \ref{halo.ps}.
Therefore, where the cold atomic gas locates is the key to understanding the
formation and quenching for red spirals.  

Actually, there have been studies on the locations of neutral atomic gas of red
spirals. \citet{Lemonias2014} provided resolved HI images observed by the
Jansky Very Large Array (VLA) for 20 HI-rich massive galaxies with low specific
star formation rates, in which there are four of our sample red spirals. They
found that the HI gas are distributed in the very extended rotational disks
with low HI surface densities, whose sizes are twice of the optical disks.
Therefore, \citet{Lemonias2014} speculated that the low specific star formation
rates may be caused by the low HI gas surface densities on the large HI disks.
More recently, \citet{Zhang2019} also found that most of the massive quenched
central disk galaxies possess a large amount of HI gas content and show
symmetric double-horn HI spectra, suggesting regularly rotating HI disks, and
the radii of HI disks are estimated to be $\sim30$ kpc according to the HI
size-mass relation. From these studies, it is clear that the cold atomic gas is
already in the extended rotating gas disks of galaxies, instead of a part of
the intergalactic medium (IGM) in
halos. This solves the puzzle of our suspicion of the halo quenching scenario.
Halo quenching is essential to preventing fresh cold gas from flowing into
galaxies. To explain why the quenched central disk galaxies have plenty of HI
gas but ceased star formation, \citet{Peng2020} proposed a new quenching
mechanism for disk galaxies -- angular momentum quenching: Once the accreted
material, mainly in the form of HI, flows in with too high angular momentum to
maintain the radial inflow, the star formation in the disk will be quenched and
the incoming gas will settle down on the outer part of the disk.  In addition,
morphological quenching may also contribute to the quenching of red spirals
given the presence of the big bulges in red spirals (Figure \ref{BThigh.ps}),
which will stabilize the gas disks, and hence star formation is prohibited
\citep{Martig2009}. To summarize, halo quenching, angular momentum quenching,
and morphological quenching may have been working together to cease the star
formation in red spirals.

\citet{Belli2019} claimed that quenching processes are closely linked with the
formation processes, i.e., the physical mechanisms responsible for quenching
are tightly related to the properties of progenitors. The different quenching
processes of ellipticals and red spirals may imply that their formation
processes are also different.  As stated in Section \ref{subsec:morphology},
massive ellipticals formed mainly via two phases
\citep[e.g.,][]{Oser2010,vanDokkum2010}, i.e., the formation of compact cores
via violent gas-rich processes, including violent disk instabilities or
gas-rich mergers, followed by the buildup of extended outer regions through
minor dry mergers.  As for red spirals, by comparing the MaNGA data analysis
for red spirals with simulations \citep{Springel05, Robertson06, Hopkins09,
Athanassoula2016, SparreSpringel2017}, \citet{Hao2019} speculated that red
spirals are results of very gas-rich major mergers at high redshift, instead of
violent disk instabilities. During such a gas-rich major merger, gas within
some characteristic radius lose angular momentum, then fall into the center
rapidly and form a compact bulge in a starburst mode. Meanwhile, the gas and
stars without losing a significant amount of angular momentum will form a
rotational disk surrounding the bulge in the merger remnant. Our results on the
stellar population properties of the central regions and the morphologies of
red spirals in this work are also coincident with this formation scenario.
Although the specific formation processes are different, both the central
regions of ellipticals and red spirals formed via very gas-rich processes that
induce starbursts in galaxy centers. The rapid gas-rich processes and the
associated starbursts could result in the close properties of the central
regions of ellipticals and red spirals in stellar populations and structures. 

Furthermore, \citet{DekelBurkert2014} pointed out that only half of the
$z\sim2$ star-forming galaxies can form compact cores and the remaining half
became extended stellar disks due to the log-normal distribution of spin
parameters. Accordingly, we suggest a scenario: The instabilities of very
gas-rich disks with low angular momentum can form compact cores of ellipticals,
while red spirals were formed by major mergers between the extended disk
galaxies with high angular momentum before most of the gas turned into stars.
Our conjecture can be tested using modern high-resolution cosmological
hydrodynamical simulations, such as IllustrisTNG
\citep{Marinacci2018,Naiman2018,Nelson2018,Pillepich2018,
Springel2018,Nelson2019,Pillepich2019}, by tracing back the formation histories
of red spirals. Before such a sophisticated work, a consistency check on the
number density of $z\sim2$ gas-rich major mergers with that of our red spirals
and high $\Sigma_1$ blue spirals would be helpful. There are several obstacles
to an accurate comparison between these number densities. For one thing, our
galaxy samples are representative but not complete. For the other, the merger
rate of high-z galaxies suffers from large uncertainties and the redshift
interval for the integration is also uncertain.  Despite these difficulties, a
rough estimate can be made. Considering the number density of $z\sim2$ massive
star-forming galaxies of $10^{-3}$\,Mpc$^{-3}$ \citep[e.g.,][]{Muzzin2013} and
the major merger fraction of main-sequence galaxies at the corresponding
redshift of 5-10\% \citep{Cibinel2019}, the number density of gas-rich major
mergers at $z\sim2$ would be $5\times10^{-5}$-$10^{-4}$\,Mpc$^{-3}$. This
agrees with the estimate of \citet{Barro2013} based on earlier studies. For our
red spirals, their number density was estimated using two methods. One is based
on the combination of the number density of massive star-forming galaxies at
$z\sim0.1$ \citep{Moustakas2013} over the mass range of our samples
($\sim7\times10^{-4}$\,Mpc$^{-3}$) and the fraction of red spirals (279/1,914).
The other is based upon the combination of the number density of $z\sim0.1$
massive quiescent galaxies ($10^{-3}$\,Mpc$^{-3}$) and the number ratio of red
spirals to red ellipticals (279/2,889). Inspiringly, these two methods produced
consistent results, with the number density of red spirals about
$10^{-4}$\,Mpc$^{-3}$.  Similarly, the number density of high $\Sigma_1$ blue
spirals is estimated to be $\sim5\times10^{-5}$\,Mpc$^{-3}$. Therefore, the
combined number density of red spirals and high $\Sigma_1$ blue spirals is
roughly consistent with the number density of $z\sim2$ gas-rich major mergers,
which lends further support to our proposed scenarios for the formation
mechanisms of red spirals and high $\Sigma_1$ blue spirals.

\section{SUMMARY}

For the purpose of unveiling the formation and quenching processes for massive
red spiral galaxies, we select a sample of massive red spiral galaxies, as well
as samples of blue spiral and red elliptical galaxies, with $M_{\ast} >
10^{10.5}M_{\odot}$ from the stellar mass catalog of \citet{Mendel2014} that is
based on the SDSS DR7 \citep{York2000, Abazajian2009}.  Under the constraint of
redshift range of $0.02<z<0.05$ and the luminosity with $M_{z,\rm Petro} <
-19.5$ mag, our massive galaxy sample consists of 279 red spirals, 961 blue
spirals and 2,889 red ellipticals, respectively. The main results are
summarized as follows.

\begin{enumerate}
\item We find that the red spirals and ellipticals are located in the same
region in the central spectral indices [Mg/Fe] vs. D$_n(4000)$ diagram, and the
relation between [Mg/Fe] and D$_n(4000)$ followed by them is similar. In
contrast, the blue spirals are located in a completely different region in this
diagram, being younger and less $\alpha$-element enhanced. Given that
D$_n(4000)$ and [Mg/Fe] are age and star formation timescale indicators for
galaxies, respectively, the similar ranges and relations of D$_n(4000)$ and
[Mg/Fe] followed by the central regions of red spirals and ellipticals suggest
their similar formation epoch and star formation timescale, i.e., by redshift
$\sim1-2$ and within $\sim$ 1\,Gyr. 

\item We also find that most red spirals harbor compact cores with high stellar
mass surface densities measured by $\Sigma_1$, and quite a large fraction of
red spirals is bulge-dominated.  In particular, the red spirals, especially the
bulges of red spirals follow the same $\Sigma_1$-$M_{\ast}$ relation for
quenched galaxies, in terms of both the slope, 1$\sigma$ vertical scatters and
the intercept. It supports the statement based on the spectroscopic analysis
that the bulges of red spirals and ellipticals formed in the same epoch and
quenched rapidly. Moreover, the cumulative halo mass distribution of central
red spirals is also similar to that of ellipticals, and most of them have halo
masses larger than $10^{12} M_{\odot}$. Therefore, halo quenching,
morphological quenching and the most recently proposed angular momentum
quenching mechanisms may jointly play a role in quenching red spirals.
Furthermore, by careful morphological examinations, we find that quite a large
fraction ($\sim 70\%$) of red spirals show abnormal morphologies with strong
bars, inner and outer rings (shells), and even tidal streams and other merger
remnants. These results are consistent with the simulations in which very
gas-rich major mergers can form disk galaxies.
 
\item The investigations for high $\Sigma_1$ blue spirals reveal their
similarities to red spirals in morphology, bulge mass, $B/T$ distribution, as
well as halo mass range and HI detection rate and mass fraction. This strongly
suggests that the high $\Sigma_1$ blue spirals have experienced similar
formation histories to red spirals. Considering the fact that there is on-going
star formation in high $\Sigma_1$ blue spirals, one of the reasonable
explanations for the similarities and dissimilarities between high $\Sigma_1$
blue spirals and red spirals would be rejuvenation from red spirals to blue
spirals. The rejuvenation is likely induced by interactions or bar
instabilities, which make cold gas flow into the galaxies and re-ignite star
formation in the bulges and disks of red spirals, driving them towards blue.

\end{enumerate}

\acknowledgments

We thank the anonymous referee for his/her constructive and helpful comments
that improved the paper.  We would like to thank Drs. Shude Mao, Cheng Li,
Chenggang Shu, Fangzhou Jiang, Jian Fu and Dandan Xu for advice and helpful
discussions. We acknowledge Drs. J. T. Mendel and L. Simard for providing the
$u$-band bulge-disk decomposition data.  This work is supported by the National
Key Research and Development Program of China (No.  2017YFA0402703) and the
National Natural Science Foundation of China (NSFC, No.  11733002 and
11373027). Y.S. also acknowledges the support from the National Key R\&D
Program  of  China  (No.  2018YFA0404502, No. 2017YFA0402704), the NSFC (No.
11825302, 11773013)  and the Excellent Youth Foundation of the Jiangsu
Scientific  Committee (BK20150014). Y.C. also acknowledges the support from the
National Key R\&D Program of China (No. 2017YFA0402704) and the NSFC (No.
11573013).  Funding for the creation and distribution of the SDSS Archive has
been provided by the Alfred P. Sloan Foundation, the Participating
Institutions, the National Aeronautics and Space Administration, the National
Science Foundation, the U.S.  Department of Energy, the Japanese
Monbukagakusho, and the Max Planck Society.  The SDSS Web site is
http://www.sdss.org/.  The SDSS is managed by the Astrophysical Research
Consortium (ARC) for the Participating Institutions. The Participating
Institutions are The University of Chicago, Fermilab, the Institute for
Advanced Study, the Japan Participation Group, The Johns Hopkins University,
the Korean Scientist Group, Los Alamos National Laboratory, the
Max-Planck-Institute for Astronomy (MPIA), the Max-Planck-Institute for
Astrophysics (MPA), New Mexico State University, University of Pittsburgh,
Princeton University, the United States Naval Observatory, and the University
of Washington.

\end{document}